\documentstyle[12pt]{article}
\def\half{\textstyle{1\over 2}}
\def\quart{\textstyle{1\over 4}}
\def\NP#1{Nucl. Phys. {\bf #1}}
\def\PL#1{Phys. Lett. {\bf #1}}

\def\NC#1{Il Nuovo Cim. {\bf #1}}

\begin{document}

\hspace{11.6cm} DFUPG 97/94

\begin{center}

\baselineskip=24pt plus 0.2pt minus 0.2pt
\lineskip=22pt plus 0.2pt minus 0.2pt

 \Large

How and why the wave function collapses after a measurement\\

\vspace*{0.35in}

\large

Massimo Fioroni$^1$  and Giorgio Immirzi$^{1,2}$ \\
\vspace*{0.25in}

\normalsize
  $^1$ \small Dipartimento di Fisica, Universit\'a di Perugia\\
$^2$\small  Istituto Nazionale di Fisica Nucleare, sezione di Perugia\\
\small  via A. Pascoli, 06100 Perugia, Italy\\
{\small\it  email: fioroni@perugia.infn.it ,
   immirzi@perugia.infn.it}\\
\vspace{.5in}
November 15, 1994\\
\vspace{.5in}
ABSTRACT
\end{center}
\vspace{.5in}
We explain the collapse of the wavefunction with the notion that, in a
measurement, the system observed nucleates a first order phase transition
in the measuring device. The possible final states differ by the values
of macroscopic observables, and their relative phase is therefore
unobservable. The process is irreversible, but needs no separate postulate.

\pagebreak

\baselineskip=24pt plus 0.2pt minus 0.2pt
\lineskip=22pt plus 0.2pt minus 0.2pt

\setcounter{page}{1}
The purpose of this paper is to restate and expand the explanation
of the collapse of the wavefunction suggested by various authors
\cite{aa,bb,cc,dd}\footnote{see \cite{aa,ee} for an adequate list of
references,
much longer than this paper.}, perhaps most clearly by K. Hepp \cite{ff}.
Briefly, this is the view that  Von Neumann's
``process 1'' \cite{gg}, which occurs in a system when a measurement is
performed, is
a short hand description of what happens when the system observed
triggers a first order phase transition in the measuring
apparatus. The possible final states of the system+detector become
macroscopically different and therefore ``disjoint'',
i.e. there is no way to measure their relative phases, hence they
are best described by a density matrix. This being the
nature of the measuring process, it is clearly irreversible,
but does not require a separate postulate.

We have found that this point of view, expressed in a variety of
ways, is informally accepted by many working physicists, and
even taken for granted by experimentalists (e.g. \cite{hh}).
Unfortunately, L.D. Landau held the opposite opinion \cite{jj},
and conjectured that on the contrary, the law of increase of
entropy might follow from the postulate that the wave function collapses, i.e.
that
process 1 occurs when a measurement is performed.

Let us briefly recall what the problem is \cite{kk}, in the idealized
description given by J. Von Neumann \cite{gg}. To measure the value
of an observable $O$, with  discrete spectrum (i.e. $\hat O\psi_n =
o_n\psi_n$, or  $\hat O = \sum o_n\hat P_n$) on a system S with
wavefunction  $\psi$, we need to couple it to  a classical
measuring apparatus M, initially  in a state $\Psi_0$, and to
arrange the interaction between the two in such a way that for each
$n$, $\psi_n\Psi_0$ quickly evolves to some $\psi_n\Psi_n$; then
inspecting the state of M we can figure out what the state of S was.
Von Neumann proves that a unitary operator realizing
this transition exists, and therefore, if S is initially in a
superposition of eigenstates of $O$, we would expect:
\begin{equation}
\sum c_n\psi_n\Psi_0  \longrightarrow \sum c_n\psi_n\Psi_n
\label{eq:i}\end{equation}
On the contrary, we find that the wave function ``collapses'',
so that after the measurement the compound system has to be described
by the density matrix
\begin{equation}
\hat\rho =
\sum |c_n|^2\;\psi_n \Psi_n\otimes\psi_n^\dagger\Psi_n^\dagger
\label{eq:ii}\end{equation}
How can this irreversible process be compatible with
the unitary evolution one expects from the Schr\" odinger
equation?  \footnote{note that in either case the
final state of the system will be described by a density matrix
if we ignore the result of the measurement, i.e. trace over the
variables of M. This is all one needs e.g. in Feynman's approach.}

Reflecting on concrete examples of measuring devices, one comes
to the conclusion that  M must be ``classical'' in the sense
that it can be, and is described by thermodynamic variables,
i.e. macroscopic order parameters defined by averages over a large
number of microscopic observables. Initially M is in a
metastable state, so that its free energy is at some local minimum.
The interaction with the microscopic system S triggers a first order
phase transition for M, to a state of lower free energy
which is a true minimum. This process is typically irreversible:
of course the microscopic dynamics is governed by the Schr\" odinger
equation, but the evolution of the order parameter(s) is
irreversible, something we have known since the work of L.
Boltzmann\footnote{a
beautiful discussion of Boltzmann's work on the origin of
irreversibility has been given by J.L. Lebowitz \cite{ll}, whose point of
view on the  collapse of the wave function is, incidentally,
completely consistent with ours.}.

If the final states correspond to different phases of M, that are
macroscopically different, i.e. differ by the value of a
macroscopic quantity, they cannot be
meaningfully superimposed, because no observation  can reveal their
relative phases, and {\it must} be treated like mixtures, as in
eq.~\ref{eq:ii}:
this basic superselection rule, which disposes of
Schr\"odinger's cat paradox, is the  point emphasized and elucidated
by K. Hepp \cite{ff}. This fact is familiar when the
different phases correspond to different directions of spontaneous
symmetry breaking \cite{mm}, but holds much more generally
as a natural consequence of the modern formulation of
quantum statistical mechanics \cite{nn}.

In this language, one takes as primary objects the
$C^*$ algebra ${\cal A}$ of the local {\it observables}, and the
{\it states}, positive linear functionals which associate to every
observable an expectation value. For every state we may find a
representation of ${\cal A}$ as an algebra of operators on a
Hilbert space. In the thermodynamic limit, e.g. when the number of
constituents $N\rightarrow\infty$, these representations may be
equivalent or not, with equivalence classes labeled by the
values of the macroscopic observables, obtained as limits from
local ones. Non equivalence means that no observable has matrix
elements between the state vectors of the two representations,
which  are therefore ``disjoint''.

For example \cite{ff}, if we label the states of a
spin$-\half$ by the unit 3--d vectors ${\bf e} =
(\sin\theta \cos\phi ,\sin\theta\sin\phi ,\cos \theta )$ such that
$(\sigma\cdot{\bf e})\,u_{\bf e}= u_{\bf e}$, we find:
\begin{equation}
u_{\bf e} =\left(\matrix{ \cos{\theta\over 2}e^{-i\phi /2}\cr
  \sin{\theta\over 2}e^{+i\phi /2}\cr}\right)\ ;\quad
|u_{{\bf e}'}^\dagger\cdot u_{\bf e}|^2={1+{\bf e}'\cdot{\bf e}\over 2}
\leq \exp (-\quart |{\bf e}'- {\bf e}|^2)
\label{eq:iii}\end{equation}
(by $1+x<e^x $). Then for the states of a system made of $N$ spin$-\half$
\begin{equation}
|<{\bf e}'_1\ldots {\bf e}'_N | {\bf e}_1\ldots {\bf e}_N>|^2 \leq
\exp (-\quart\sum |{\bf e}'_k-{\bf e}_k|^2)\leq
\exp (-\quart N | {1\over N}\sum {\bf e}'_k
-{1\over N} \sum {\bf e}_k |^2)
\label{eq:iv}\end{equation}
The same bound will drive to zero the matrix elements of any
operator involving a finite number of spins;
therefore, in the thermodynamic limit $N\rightarrow \infty $,
no physical measurement can measure  the relative phase between
states which differ by the mean value of the spin vector
${1\over N}\sum{\bf e}_k$, because no operator can change all
the spins simultaneously.

There are  other reasons that make the idea that quantum
measurements happen through first order phase transitions
in the measuring device a rather attractive one. Very loosely speaking, one may
wonder in general how can a ``small'' system affect a ``large'' one. But
typically metastability is  precipitated by a microscopic
nucleation mechanism, which only requires microscopic energies,
although the energy released in the transition is certainly
macroscopic. In this respect simple explicit models (e.g. \cite{oo,ff,dd})
which simple or no  internal dynamics of M, are inevitably inadequate
\cite{pp,qq,ee}.
On the other hand nucleation theory has a long history, from Van der Waals,
to Becker and D\"oring, to a vast modern literature. A well studied
theoretical model \cite{rr} is the two dimensional Ising model with
\begin{equation}
-\beta H = b\sum_{<ij>}\sigma_i\sigma_j + h\sum_i\sigma_i
\label{eq:v}\end{equation}
taken below  the Curie point $b=0.44..$ with, say,
positive  magnetization $<\!\sigma\!>$, and a small magnetic field in the
opposite
direction. A finite system like this is not at equilibrium, but
actually quite stable, unless a ``droplet'' of spins is inverted.
As an example, we run a Montecarlo simulation on a 200x200
lattice, thermalizing it with 3000 runs at $h=0,\ b=0.6$, then
going slowly to $h=-0.05$. No inversion of $<\!\sigma\!>$ takes place
for 9000 runs, unless a square of at least 12x12 spins is inverted,
whereupon $<\!\sigma\!>$ changes sign within a few hundred runs. Of course this
is a classical model, but one does not expect a quantum model
to behave very differently.

It may be that the explanation we give makes the wave
function collapse rather trivial; and indeed, it is difficult
to justify the interest and the immense literature the matter has
attracted over the years, to say nothing about the extravagant
philosophical implications (partly listed in \cite{dd}).
Let us emphasize that in our opinion the wave function collapse does
not play an important role in the theory or the practice of quantum mechanics,
and  has nothing directly to do with the profound and important
epistemological questions that arise because of the probabilistic
nature of quantum mechanics, nor with the possibility of
recostructing a consistent history of a phenomenon from a sequence
of observations. An early paper by N.F. Mott \cite{ss} is particularly
illuminating on this last subject: the straight tracks produced in a
Wilson chamber by an $\alpha$ particle emitted in S-wave follow from
the application of the Schr\"odinger equation to the combined system
$\alpha$-particle+atoms, without resorting to Von Neumann's
process 1.

On the other hand, it may be that the notion of ``disjointness'' we
have put at the centre of our analysis has a broader meaning than
the one given here. For example, it appears that the detection of a
particle by a grain of emulsion is a  more subtle and complex
phenomenon than commonly realized. A more extravagant example could
be a measuring device which worked permuting the sequence of bases
of a DNA molecule. In such cases, the possible final states of the
measuring device cannot be said to be different phases of a macroscopic
system, although it is doubtful that they can be meaningfully superimposed.
A similar question has been discussed in \cite{tt} in a simpler context, while
A.J. Leggett has proposed a quantitative measure of
"disjonctivity" \cite{uu}. On balance, we cannot consider settled the question
of wave
function collapse.

We would like to thank  M. Cini and M. Pascazio for some very helpful
discussion.

\end{document}